\documentclass[]{spie}  

\usepackage{amsmath,amsfonts,amssymb}
\usepackage{graphicx}
\usepackage{hyperref}

\begin{document} 

\title{SPHERE IRDIS and IFS astrometric strategy and calibration}

\author{Anne-Lise~Maire\supit{a},\,Maud~Langlois\supit{b,c},\,Kjetil~Dohlen\supit{c}, Anne-Marie~Lagrange\supit{d}, Raffaele~Gratton\supit{e}, Ga\"el Chauvin\supit{d}, Silvano Desidera\supit{e}, Julien H. Girard\supit{f,d}, Julien Milli\supit{f,d}, Arthur Vigan\supit{c}, G\'erard Zins\supit{f}, Philippe Delorme\supit{d}, Jean-Luc Beuzit\supit{d}, Riccardo U. Claudi\supit{e}, Markus Feldt\supit{a}, David Mouillet\supit{d}, Pascal Puget\supit{d}, Massimo Turatto\supit{e}, Fran\c cois Wildi\supit{g}
\skiplinehalf
\supit{a}Max Planck Institute for Astronomy, K\"onigstuhl 17, 69117 Heidelberg, Germany \\
\supit{b}CRAL, Univ. Lyon 1, ENS Lyon, CNRS, 9 av. Ch. Andr\'e, 69561 Saint-Genis-Laval, France \\ 
\supit{c}LAM, Aix Marseille Univ., CNRS, B.P. 8, 13376 Marseille, France \\
\supit{d}IPAG, Univ. Grenoble Alpes, CNRS, B.P. 53, 38041 Grenoble, France \\
\supit{e}INAF, Astronomical Observatory of Padova, Vicolo dell'Osservatorio 5, 35122 Padova, Italy \\
\supit{f}ESO, Alonso de Cordova 3107, Casilla 19001 Vitacura, Santiago 19, Chile \\
\supit{g}Geneva Observatory, Univ. of Geneva, Chemin des Maillettes 51, 1290 Versoix, Switzerland \\
}

\authorinfo{Further author information: send correspondence to maire@mpia.de.}

\pagestyle{empty} 
\setcounter{page}{301} 

\maketitle

\begin{abstract}
We present the current results of the astrometric characterization of the VLT planet finder SPHERE over 2 years of on-sky operations. We first describe the criteria for the selection of the astrometric fields used for calibrating the science data: binaries, multiple systems, and stellar clusters. The analysis includes measurements of the pixel scale and the position angle with respect to the North for both near-infrared subsystems, the camera IRDIS and the integral field spectrometer IFS, as well as the distortion for the IRDIS camera. The IRDIS distortion is shown to be dominated by an anamorphism of 0.60$\pm$0.02\% between the horizontal and vertical directions of the detector, i.e. 6~mas at 1~arcsec. The anamorphism is produced by the cylindrical mirrors in the common path structure hence common to all three SPHERE science subsystems (IRDIS, IFS, and ZIMPOL), except for the relative orientation of their field of view. The current estimates of the pixel scale and North angle for IRDIS are 12.255$\pm$0.009~milliarcseconds/pixel for H2 coronagraphic images and -1.75$\pm$0.08$^{\circ}$. Analyses of the IFS data indicate a pixel scale of 7.46$\pm$0.02~milliarcseconds/pixel and a North angle of -102.18$\pm$0.13$^{\circ}$. We finally discuss plans for providing astrometric calibration to the SPHERE users outside the instrument consortium.

\end{abstract}


\section{INTRODUCTION}
\label{sec:intro}  
The detection and characterization of exoplanets is one of the most active research areas in modern astrophysics. Many methods are employed to address specific types of objects and/or questions. High-contrast imaging is currently the most efficient technique for probing (1) the architecture of systems with wide-separated (beyond $\sim$5--10 AU) Jovian mass planets and (2) the spectral properties of these planets\cite{Marois2008c, Lagrange2010b, Macintosh2015}\,. A key science motivation for
these studies is to understand whether this population of planets is the large-separation tail of the distribution
of planets discovered by radial velocity and transit surveys at shorter separations or they form a separated population with different formation mechanisms. Astrometry is critical to confirm the companionship of detected companion candidates from the comparison of the measured relative position at two separated epochs to the predicted positions under the hypothesis that they are background objects. For short-period systems, accurate astrometric monitoring is needed to constrain the orbital elements of the individual companions (e.g., period, semi-major axis, eccentricity, inclination) and the total dynamical mass hence provides insights into the dynamical properties (e.g., mean motion resonances, dynamical interactions, orbital stability). The orbital elements can be compared to predictions from different formation scenarios for substellar companions (core accretion, gravitational instability in a circumstellar disk, collapse and fragmentation of a molecular cloud) to constrain the formation mechanisms of the systems\cite{Bate2009, Raghavan2010}\,. If radial velocity measurements or astrometric motion of the central star are available, the dynamical mass of the companions can be constrained\cite{Close2005, Crepp2012, Bonnefoy2014c}\,. Dynamical mass measurements provide critical tests for atmospheric and evolutionary models for brown dwarfs and giant planets\cite{Burrows1997, Baraffe2003, 2012RSPTA.370.2765A, Spiegel2012}\,. Most of the young directly-imaged substellar companions have semi-major axis larger than a few tens of AU, so that estimates for their mass can only be inferred from their luminosity assuming possible ranges for the system age and evolutionary models. However, these estimates are strongly uncertain because the models are poorly calibrated at young ages and low masses. Detecting and measuring the orbit of young and low-mass substellar companions provide valuable benchmarks for the models.

The detection and the orbital analysis of young and low-mass substellar companions are a major part of the science case for the recently commissioned planet finder SPHERE\cite{Beuzit2008} (Spectro-Polarimetric Exoplanet REsearch). The instrument includes an extreme adaptive optics system\cite{Fusco2014}\,, with a pupil stabilization control system and stress polished toric mirrors\cite{Hugot2012} to relay the beam to the coronagraphs\cite{Boccaletti2008c} and the science instruments. The science instruments are composed of the infrared dual-band imager and spectrograph IRDIS\cite{Dohlen2008a}\,, the near-infrared integral field spectrometer IFS\cite{Claudi2008}\,, and the rapid-switching visible imaging polarimeter ZIMPOL\cite{Thalmann2008}\,. SPHERE has been successfully commissioned at the Very Large Telescope from May to October 2014 and is offered to the community since April 2015. The SPHERE consortium guaranteed-time survey consists of 260 nights over 5 years, from which 200 nights are dedicated to a large census in the near-infrared of the population of young giant planets and brown dwarfs at wide orbits ($\gtrsim$5~AU). The main observing mode used for this survey consists in simultaneous observations in the YJ bands (0.95--1.35~$\mu$m, $R$\,$\sim$\,54) with IFS (field of view 1.73$''$\,$\times$1.73$''$) and in the H-band (H$_2$\,=\,1.593~$\mu$m and H$_3$\,=\,1.667~$\mu$m) with IRDIS (field of view 11$''$\,$\times$12.5$''$) in dual-band imaging mode\cite{Vigan2010}\,. A coronagraphic mask common to both instruments is used to attenuate the stellar light\cite{Boccaletti2008c}\,. Both IRDIS and IFS are operated in pupil-stabilized mode to take advantage of the angular differential imaging technique\cite{Marois2006a} to further suppress the stellar residuals in the images. High-precision relative astrometry and efficient attenuation of the stellar residuals by image post-processing techniques critically depend of a precise estimate of the location of the star behind a coronagraphic mask\cite{Marois2006b, Sivaramakrishnan2006}\,. For this purpose, a calibration image is recorded before and after a science sequence with four crosswise faint stellar replicas produced by applying a periodic modulation on the SPHERE deformable mirror\cite{Langlois2013}\,. The specifications for the SPHERE astrometric accuracy are 5 mas (goal 1 mas). Extensive tests using injections of synthetic point sources in laboratory data processed with spectral differential imaging\cite{Racine1999, Sparks2002} resulted in astrometric accuracies below 1.5--2~mas for detections at signal-to-noise ratios above 10\cite{Zurlo2014}\,.

We present in this paper the current on-sky status and results of the astrometric characterization of IRDIS and IFS based on 2~years of SPHERE operations from 2014 to 2016. We describe in Sec.~\ref{sec:methodology} our criteria for the selection of the astrometric fields. We present in Sec.~\ref{sec:results} the on-sky measurements of the SPHERE optical distortion, the zeropoint angle of the SPHERE pupil in pupil-stabilized mode, and the pixel scale and true North offset for both IRDIS and IFS. We summarize our main results and briefly discuss a few prospects in Sec.~\ref{sec:conclusions}.

\section{METHODOLOGY}
\label{sec:methodology}

\subsection{Strategy for the SPHERE astrometric characterization}
In the case of the SPHERE planet-search observations, five parameters need to be estimated: the IRDIS pixel scale and true North offset, the IFS pixel scale and its relative orientation with respect to IRDIS, and the zeropoint angle of the SPHERE pupil in pupil-stabilized mode. The pixel scale of IRDIS is slightly dependent on the selected filter, while both IRDIS and IFS pixel scales depend on the use or not of a coronagraph because of the coronagraphic glass plate located at the focal plane. The true North offset shall be monitored regularly with time since it is expected to exhibit small variations ($<$1$^{\circ}$), especially after technical interventions on the instrument. The zeropoint angle of the instrument pupil in pupil-stabilized mode is expected to be affected by systematics due to the limited accuracy of the derotator positioning.

For the estimation of the pixel scales and the IRDIS true North offset, an absolute weekly calibration is needed using astrometric fields with accurate positions (including, if possible, the stellar proper motions). Stellar clusters are suitable targets for this purpose as they can provide several tens of stars for the analysis. Stellar cluster fields can also be used to derive the SPHERE+VLT optical distortion, although the optical distortion of the VLT is expected to be small (on-sky measurements of the Galactic Center with NaCo indicated distortion effects below 0.1 mas over a 5.4$''$ field of view\cite{Trippe2008}). Another advantage of this method over the use of internal distortion grids for measuring the SPHERE distortion is that it does not rely on a model for the reference grid. The zeropoint angles for IFS and the SPHERE pupil in pupil-stabilized mode are relative parameters, so even an astrometric binary with large uncertainties on its orbital properties is suitable for their estimation.

\subsection{SELECTION OF THE CALIBRATION FIELDS}
\begin{figure}[t]
   \begin{center}
   \begin{tabular}{c} 
   
   \includegraphics[height=7.7cm]{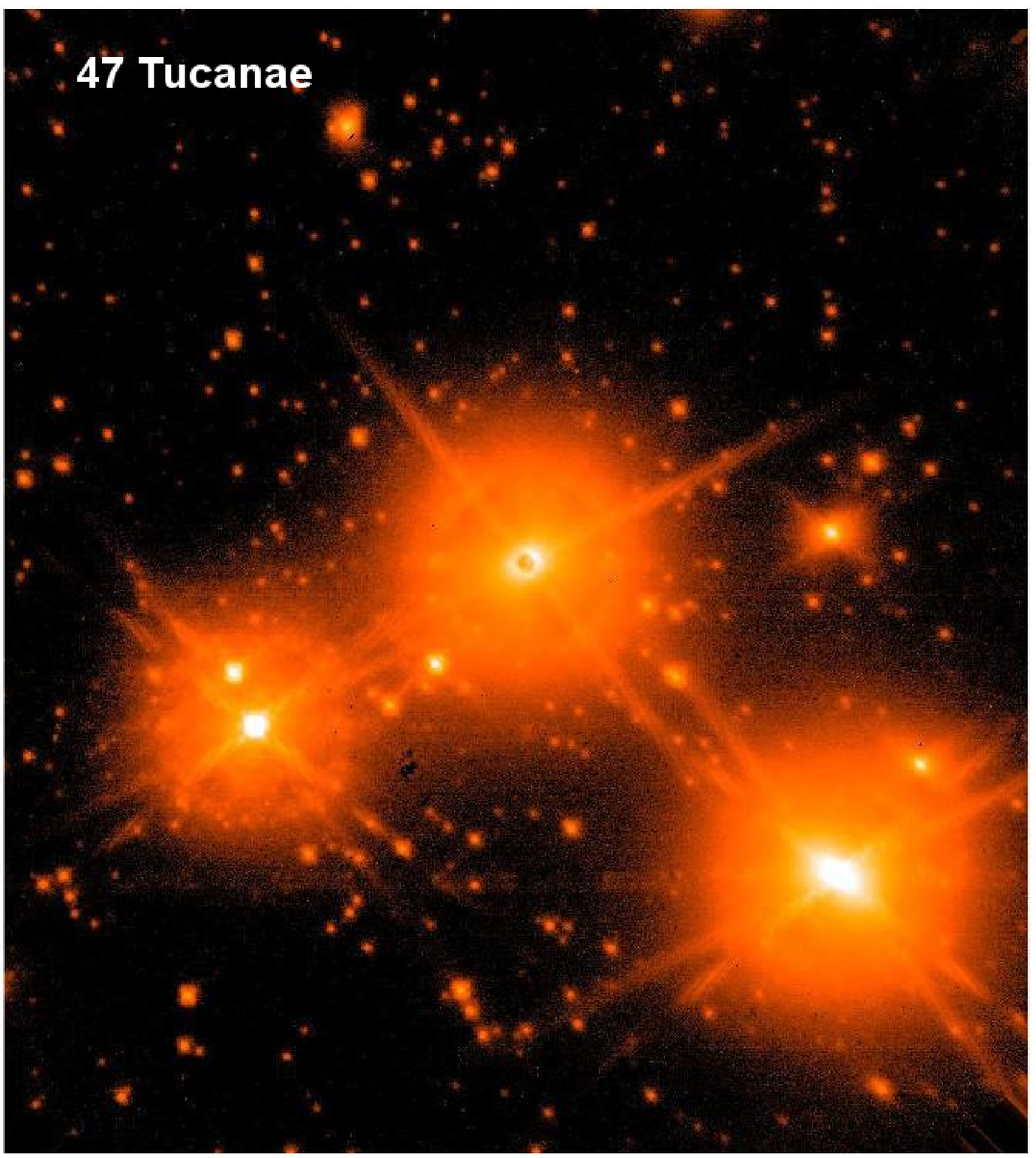}
   \includegraphics[trim = 16mm 0mm 6mm 0.1mm,clip,height=7.71cm]{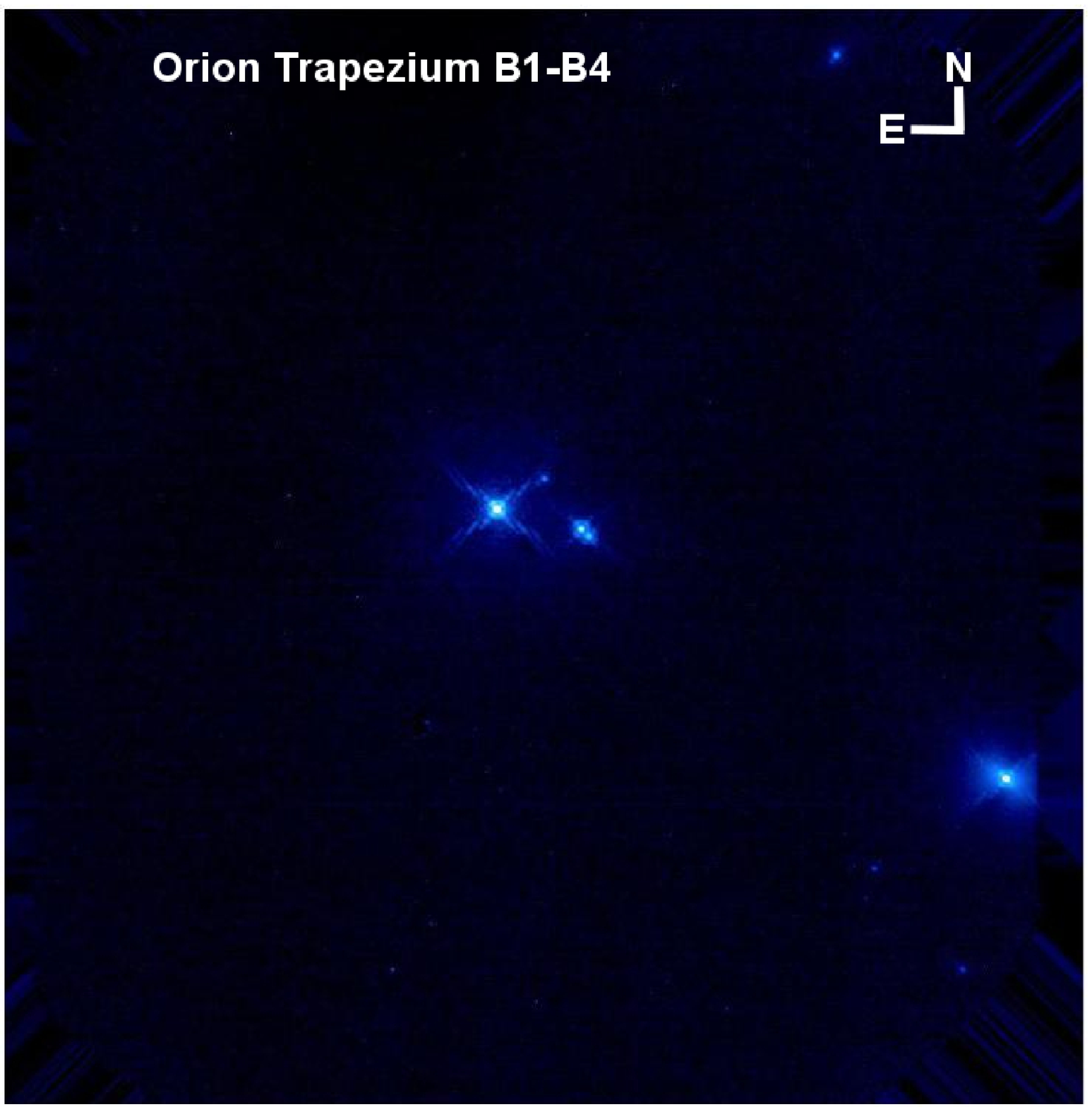} \\
    \includegraphics[height=7.7cm]{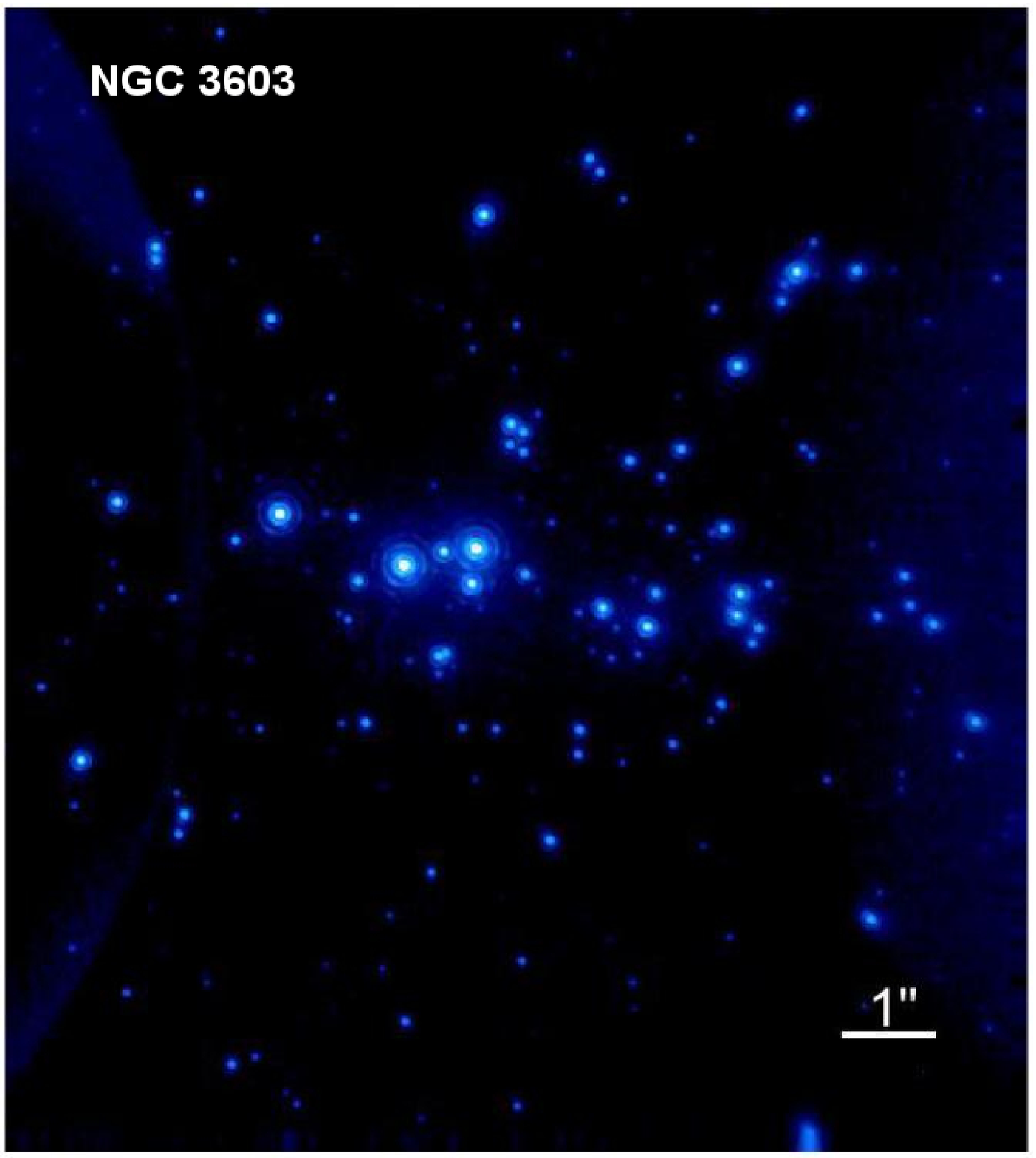}
   \includegraphics[trim = 11mm 0mm 11mm 0mm,clip,height=7.7cm]{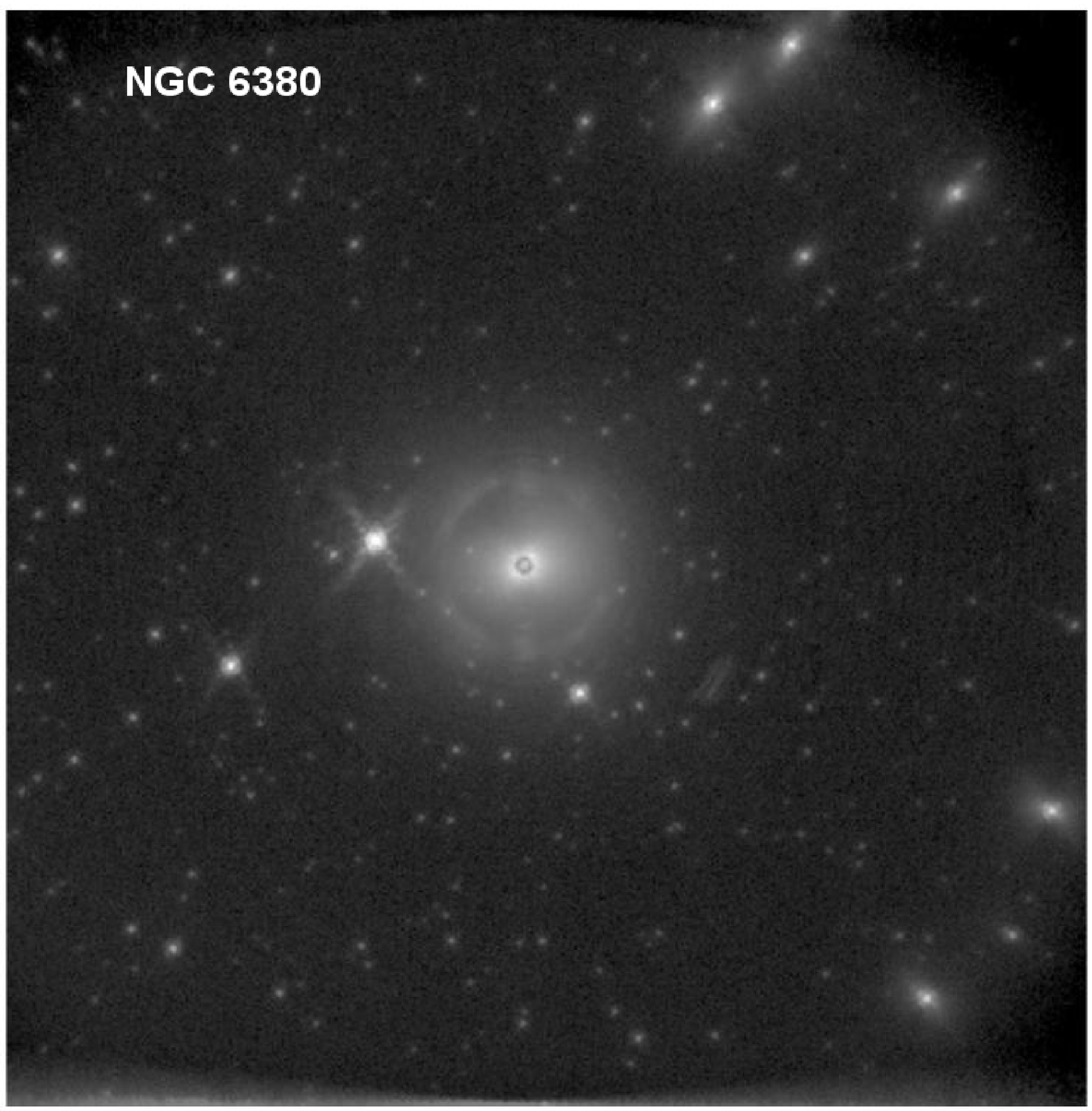}
   \end{tabular}
   \end{center}
   \caption[fields] 
   { \label{fig:fields} 
IRDIS images of the main reference SPHERE astrometric fields.}
   \end{figure}

The selection of the calibration fields was done considering the following criteria:
\begin{itemize}
\item the type of parameters to be estimated (absolute/relative); \item the availability of accurate reference positions from HST measurements;
\item the presence of a bright star for AO guiding ($R$\,$\lesssim$\,13.5~mag);
\item for binaries, an H-band magnitude for the components fainter than 8.6 mag to avoid the use of neutral
density filters to get unsaturated images at the shortest integration times (0.83\,s for IRDIS, 1.66\,s for IFS);
\item the use of the field for previous astrometric monitorings\footnote{It has not been possible to use NaCo observations for the astrometric calibration of SPHERE until
December 2015 because NaCo was removed from the VLT in September 2013 for technical interventions. After a recommissioning
on another VLT telescope in January 2015, it was again removed in March 2015 because of technical issues and was
unavailable until December 2015.};
\item a good on-sky coverage throughout the year.
\end{itemize}
We selected as main observing fields 47~Tucanae, the Orion Trapezium B1--B4, NGC~3603, and NGC~6380 (Tab.~\ref{tab:coordfields} and Fig.~\ref{fig:fields}), as well as a handful of long-period binaries (HIP~67745, HIP~68725, HIP~102979) with Hipparcos data and separations
larger than the IFS field of view (separations $\sim$2--5$''$). The catalogs of the stellar positions for these fields come from various sources: Washington Double Star (WDS) catalog\footnote{\url{http://www.usno.navy.mil/USNO/astrometry/optical-IR-prod/wds/WDS/}.} for the binaries, literature for the Orion Trapezium B1--B4\cite{Close2012}\,, and private requests for 47~Tucanae (A. Bellini/STScI), NGC~6380 (E. Noyola/Univ. Texas Austin), and NGC~3603\cite{Khorrami2016} (Z. Khorrami/OCA).

For the binaries, we fitted linear equations to the WDS data with increased weights on the Hipparcos data
for predicting their relative position with their uncertainty at the epoch of the SPHERE observations. However, the
discrepancies between the measured and expected separations and position angles are usually larger than the fitting
uncertainties. When a stellar cluster field was observed close in time to a binary, we recalibrated the predicted
position to the position derived using the astrometric calibration provided by the stellar cluster. The upcoming release of the first GAIA data in autumn 2016 will finally allow for a full determination of the absolute positions of several of our astrometric standards, allowing in the future for a more accurate absolute astrometric calibration of the SPHERE data.

When observing the Orion Trapezium B1--B4, NGC 3603 and the binaries with a coronagraph, the AO guide star is
offsetted out of the coronagraphic mask using a tip-tilt mirror so that it can be used for the astrometry. Most astrometric data were recorded in field-stabilized mode.

\begin{table}[ht]
\caption{Coordinates of the AO guide star for the stellar clusters used as references for the SPHERE astrometric fields.} 
\label{tab:coordfields}
\begin{center}       
\begin{tabular}{|l|c|c|c|} 
\hline
\rule[-1ex]{0pt}{3.5ex}  Field & 47~Tucanae & NGC~3603 & NGC~6380 \\
\hline
\rule[-1ex]{0pt}{3.5ex}  RA (J2000) & 00:23:58.12 & 11:15:07.48 & 17:34:28.11 \\
 \hline
\rule[-1ex]{0pt}{3.5ex}  DEC (J2000) & -72:05:30.19 & -61:15:38.70 & -39:04:50.56 \\
 \hline
\end{tabular}
\end{center}
\end{table}

We selected 47~Tucanae as reference astrometric field for the recalibration of the other SPHERE astrometric fields because of the availability of the individual stellar proper motions in the HST catalog (accuracy 0.3~mas/yr, reference epoch 2006.20, see [\citenum{Bellini2014}] for the methods used for their derivation).     

\section{ASTROMETRIC RESULTS FROM JULY 2014 TO MAY 2016}
 \label{sec:results}
All data were reduced with the SPHERE data reduction pipeline\cite{Pavlov2008} and analyzed with custom IDL routines to derive the astrometric calibration. For data obtained in field-stabilized mode, the individual frames are first selected based on the flux statistics and combined to enhance the signal-to-noise ratio of the detected stars. Then, the positions of the stars are measured with Gaussian fitting using the mpfit library\cite{Markwardt2009}\,. The counter-identification between the SPHERE positions and the catalog positions is done using estimates for the separations and the position angles assuming approximate values for the SPHERE pixel scale and true North offset (plus the IFS angle offset relative to IRDIS for IFS observations) and tolerance criteria. After the counter-identification, the average pixel scale and true North offset are derived from the statistics of all the available stellar pairs after removing outliers using sigma clipping. If several tens of stars are available, the SPHERE positions are corrected for the on-sky optical distortion (Sec.~\ref{sec:distortion}) before deriving the astrometric calibration. The on-sky optical distortion is measured by fitting linear coordinate transformations between the catalog and the SPHERE positions. If few stars are detected in the field, a default distortion correction is applied instead. In the case of pupil-stabilized data, the whole procedure is similar, except that the individual frames are corrected for the SPHERE distortion and derotated with respect to the first frame in the sequence prior to their combination. In this case, the estimated angle offset is the sum of the true North and the SPHERE pupil angle offsets. All angles provided in the following sections are counted positive from North to East. Work is on-going for implementing generic astrometric calibration values in the SPHERE data reduction pipeline. For now, proper astrometric calibration of the consortium guaranteed-time data is accounted for in custom IDL routines implemented in the SPHERE Data Center (IPAG/CNRS).
 
\begin{figure}[t]
   \begin{center}
   \begin{tabular}{c} 
   \includegraphics[trim = 114mm 30mm 314mm 5mm,clip,height=6cm]{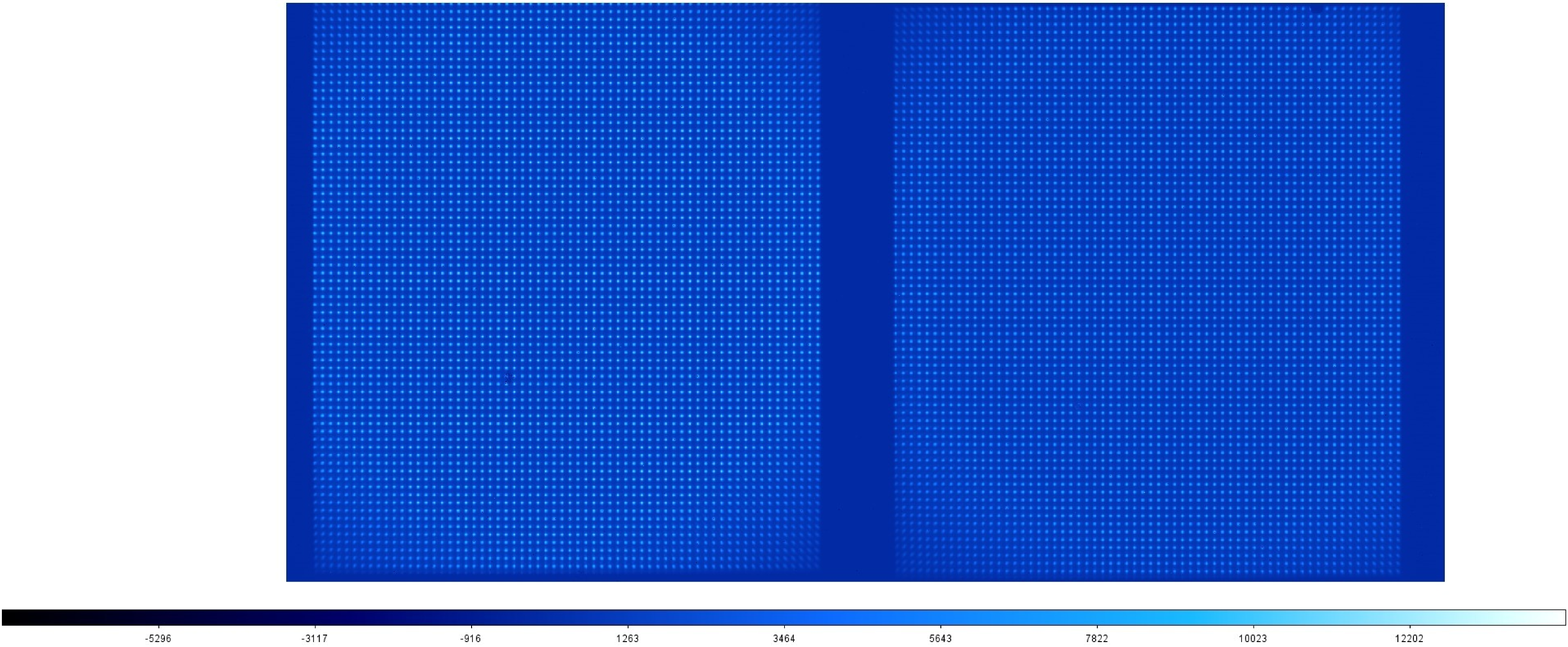}
   \includegraphics[trim = 4.8mm 10mm 4.8mm 10mm,clip,height=6cm]{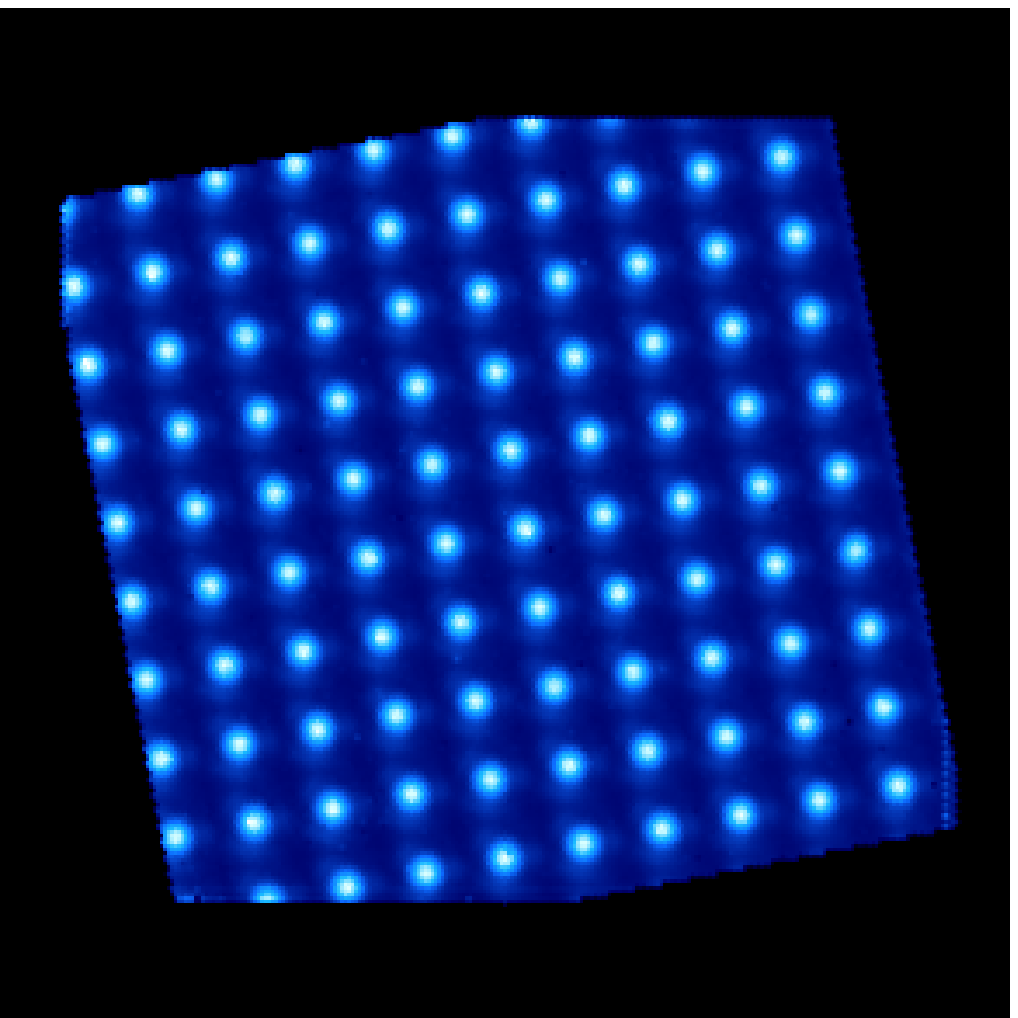}
   \end{tabular}
   \end{center}
   \caption[fields] 
   { \label{fig:distortiongrids}
   Left-side IRDIS image (left) and combined IFS image (right) of the illuminated SPHERE pinhole grid (spacing 100.0$\pm$0.5~$\mu$m). The numbers of visible pinholes differ due to different sizes of the IRDIS and IFS fields of view.}
\end{figure}
 
\subsection{SPHERE OPTICAL DISTORTION}
\label{sec:distortion}

The strategy for the measurement of the instrument optical distortion is to use (1) crowded stellar cluster fields for the absolute calibration of one instrument setup (filter pair, coronagraph) and (2) data of internal distortion grids for differential correction between different instrument setups.

A square grid of transparent dots engraved in a layer of black chrome is located in the calibration unit of SPHERE\cite{Wildi2009} hence is common to all SPHERE science subsystems (Fig.~\ref{fig:distortiongrids}). The pitch of the grid is 100.0$\pm$0.5~$\mu$m for a size of the dots of 30~$\mu$m.

Laboratory measurements showed that the cylindrical mirrors in the SPHERE common path are the main source for the instrument optical distortion, hence these distortion effects are common to IRDIS and IFS. For IFS, the distortion pattern is rotated by +100.48$\pm$0.10$^{\circ}$, the position angle of the IFS field of view with respect to the IRDIS field of view (Sec.~\ref{sec:ifs}). The distortion pattern is not affected by the stabilization mode of SPHERE (stabilization of the instrument pupil or of the on-sky field) since the instrument field derotator is the first element in the optical train.

On-sky measurements of the optical distortion using IRDIS data from 47~Tucanae and NGC~3603 confirm that the SPHERE common path optical distortion dominates the telescope distortion effects. The SPHERE distortion at first order is an anamorphism of 0.60$\pm$0.02\% between the horizontal and vertical pixel scales. This effect if uncorrected translates into an astrometric error of 6~mas at 1$''$, which is larger than the astrometric specifications for SPHERE (5~mas). Up to now, the consortium guaranteed-time data reduced at the SPHERE Data Center are corrected for this anamorphism (raw IRDIS images are vertically stretched by a factor 1.006), unlike for smaller higher-order distortion terms. The averaged error on the stellar positions over the full IRDIS  field of view when neglecting higher-order distortion terms has been measured using 47~Tucanae data and estimated to be inferior to about 1~mas, i.e. inferior to 0.09~pixel.

\subsection{ZEROPOINT ANGLE OF THE SPHERE PUPIL IN PUPIL-STABILIZED MODE}
\begin{figure}[t]
   \begin{center}
   \begin{tabular}{c} 
	 \includegraphics[height=5.8cm]{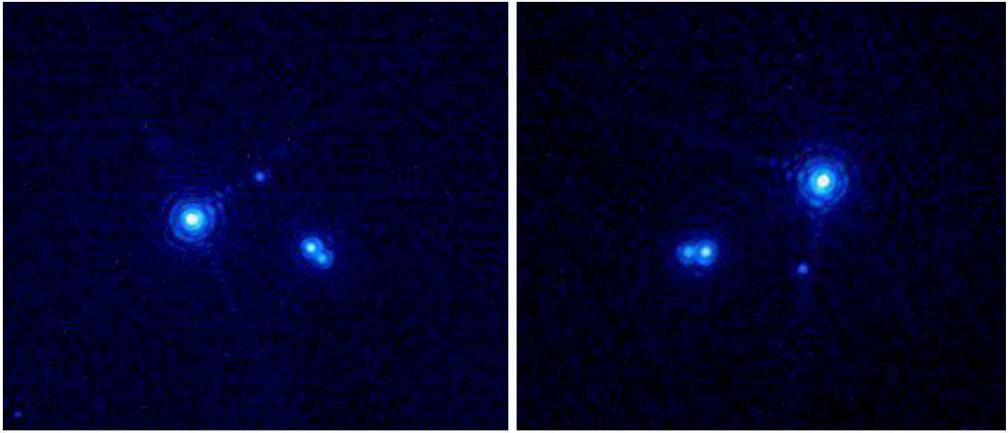}
   \end{tabular}
   \end{center}
   \caption[fields] 
   { \label{fig:fieldpuptracking} 
IRDIS images of the Trapezium B1--B4 in field-stabilized mode (left) and pupil-stabilized mode (right) for the same position angle of the SPHERE derotator.} 
   \end{figure}

To align North up and East to the left, SPHERE images obtained in pupil-stabilized mode shall also be derotated from the zeropoint angle of the instrument pupil in this mode (Fig.~\ref{fig:fieldpuptracking}). This parameter has been measured using commissioning and guaranteed-time data of several fields observed subsequently in field-stabilized mode and pupil-stabilized mode (with similar pointing parameters). The measured values are stable around an average value of $-$135.99$\pm$0.11$^{\circ}$.

\subsection{IRDIS}

\subsubsection{Pixel scale}
The pixel scale of IRDIS has been measured at several epochs for various instrument setups (filter pair, coronagraph). The apodized pupil Lyot coronagraphs and four-quadrant phase masks of SPHERE\cite{Boccaletti2008c} are made of material deposited on transparent substrates so that the pixel scale is expected to slightly change from one device to another. SPHERE also includes classical Lyot coronagraphs\cite{Boccaletti2008c}, for which the pixel scale is identical to the pixel scale measured without coronagraph (the opaque masks are suspended in this case). For given instrument setup and calibration field, measurements close in time are stable at a level of 0.009~mas/pixel. We show in the left panel of Fig.~\ref{fig:irdissumaryplots} individual pixel scale measurements for three fields observed with the H2 filter and the N\_ALC\_YJH\_S apodized pupil Lyot coronagraph (diameter 185~mas), which is the coronagraph used for the consortium guaranteed-time survey. The error bars can vary between measurements on a given stellar cluster field according to the number of stars used for the analysis (differences in integration times, observing conditions, etc). Table~\ref{tab:irdisscale} summarizes the current pixel scale estimates for almost all IRDIS filters (except for BB\_Y filter) for the N\_ALC\_YJH\_S apodized pupil Lyot coronagraph. We considered 47 Tucanae as the reference field. For the filters with no available observations with 47 Tucanae, we accounted for pixel scale systematics between the observed field and 47 Tucanae by using pixel scale measurements from the two fields obtained with the H2 filter with coronagraph. The current estimate for the pixel scale for the H2 filter with the N\_ALC\_YJH\_S coronagraph is 12.255$\pm$0.009~mas/pix (specifications 12.25$\pm$0.01~mas/pix). We note a decreasing trend for the pixel scale with the central wavelength of the filter up to the H band, followed by an increasing trend for longer wavelengths. When using no coronagraph, the pixel scale for a given filter is slightly decreased by a factor 1.0015. A calibration table with the pixel scale values for all IRDIS filter and coronagraph configurations is implemented in the SPHERE Data Center pipeline for the analysis of the consortium guaranteed-time data.

\begin{table}[ht]
\caption{On-sky measurements of the IRDIS pixel scale as a function of the filter.} 
\label{tab:irdisscale}
\begin{center}       
\begin{tabular}{|p{1.28cm}|l|l|l|l|l|l|l|l|l|l|l|l|} 
\hline
\rule[-1ex]{0pt}{3.5ex}  Filter & Y2 & Y3 & J2 & J3 & H2 & H3 & K1 & K2 & BB\_J & BB\_H & BB\_Ks\\
\hline
\rule[-1ex]{0pt}{3.5ex}  Scale (mas/pix) & 12.283 & 12.283 & 12.266 & 12.261 & 12.255 & 12.250 & 12.267 & 12.263 & 12.263 & 12.251 & 12.265 \\
 \hline
\end{tabular}
\end{center}
\end{table}

\subsubsection{True North offset}
\label{sec:irdistn}

\begin{figure}[h]
   \begin{center}
   \begin{tabular}{c} 
	 \includegraphics[width=0.49\textwidth]{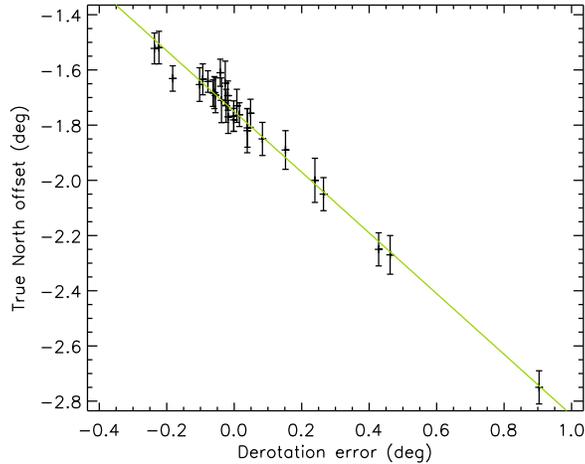}
   \end{tabular}
   \end{center}
   \caption[fields] 
   { \label{fig:tnderoterr} 
True North offset estimate as a function of the derotation error caused by missynchronization issues between the SPHERE and VLT internal clocks. A clear correlation (average fitting residuals 0.03$^{\circ}$) is seen between large true North deviations and large derotation errors.} 
   \end{figure}

The true North offset does not depend on the filter for a given observing run. We measured the true North offset at multiple epochs over 2 years of SPHERE operations (see Tab.~\ref{tab:irdistn}). While these measurements were stable within $\sim$0.15$^{\circ}$ during the commissioning and the first months of the consortium guaranteed-time survey, they exhibited anomalous variations larger than 1$^{\circ}$ between December 2015 and February 2016. Dome-tracking tests with the internal distortion grid of SPHERE showed synchronization issues between the SPHERE and VLT internal clocks. These issues were in fact present since the first light of SPHERE but somewhat mitigated by more frequent instrument resets. The left panel of Fig.~\ref{fig:tnderoterr} shows a clear correlation between the true North offset and the derotation error due to the clock missynchronization issues (average fitting residuals 0.03$^{\circ}$). The derotation error can be derived from the information in the data headers using the formula:
\begin{equation}
\epsilon\,=\,atan \left( tan \left( (ALT_{START}-PARANG_{START}-2 \times DROT2_{BEGIN}) \times \frac{\pi}{180} \right) \right) \times \frac{180}{\pi}
\end{equation}
where atan and tan ensure that $\epsilon$ is within a 360$^{\circ}$ range, $ALT_{START}$ and $PARANG_{START}$ are the telescope altitude and parallactic angles at the beginning of the observations provided by the telescope control software, and $DROT2_{BEGIN}$ is the position angle of the SPHERE derotator at the beginning of the observations calculated by the SPHERE lighting control unit (LCU).

\begin{figure}[t]
   \begin{center}
   \begin{tabular}{c} 
   \includegraphics[height=6.25cm]{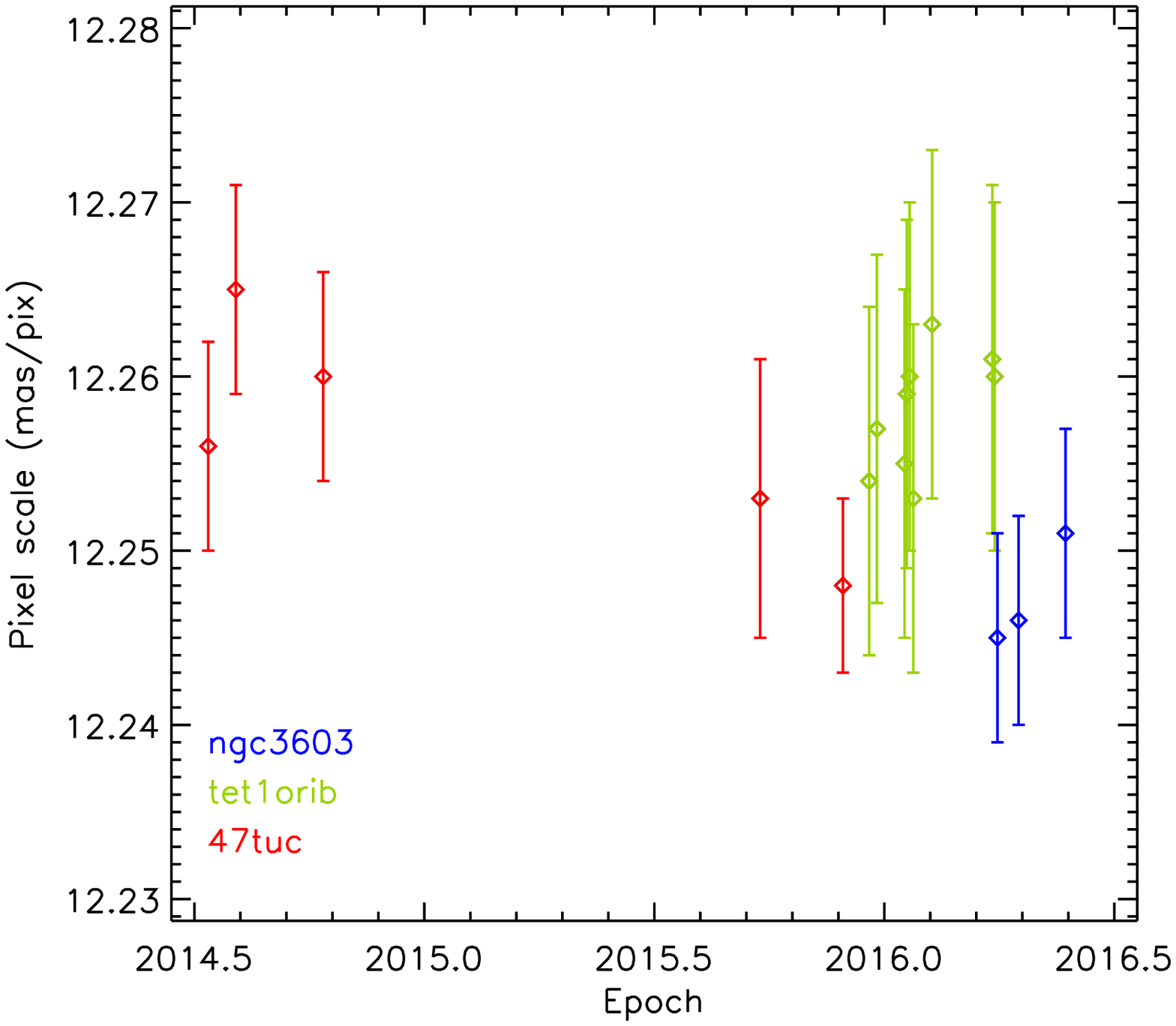}
   \includegraphics[height=6.25cm]{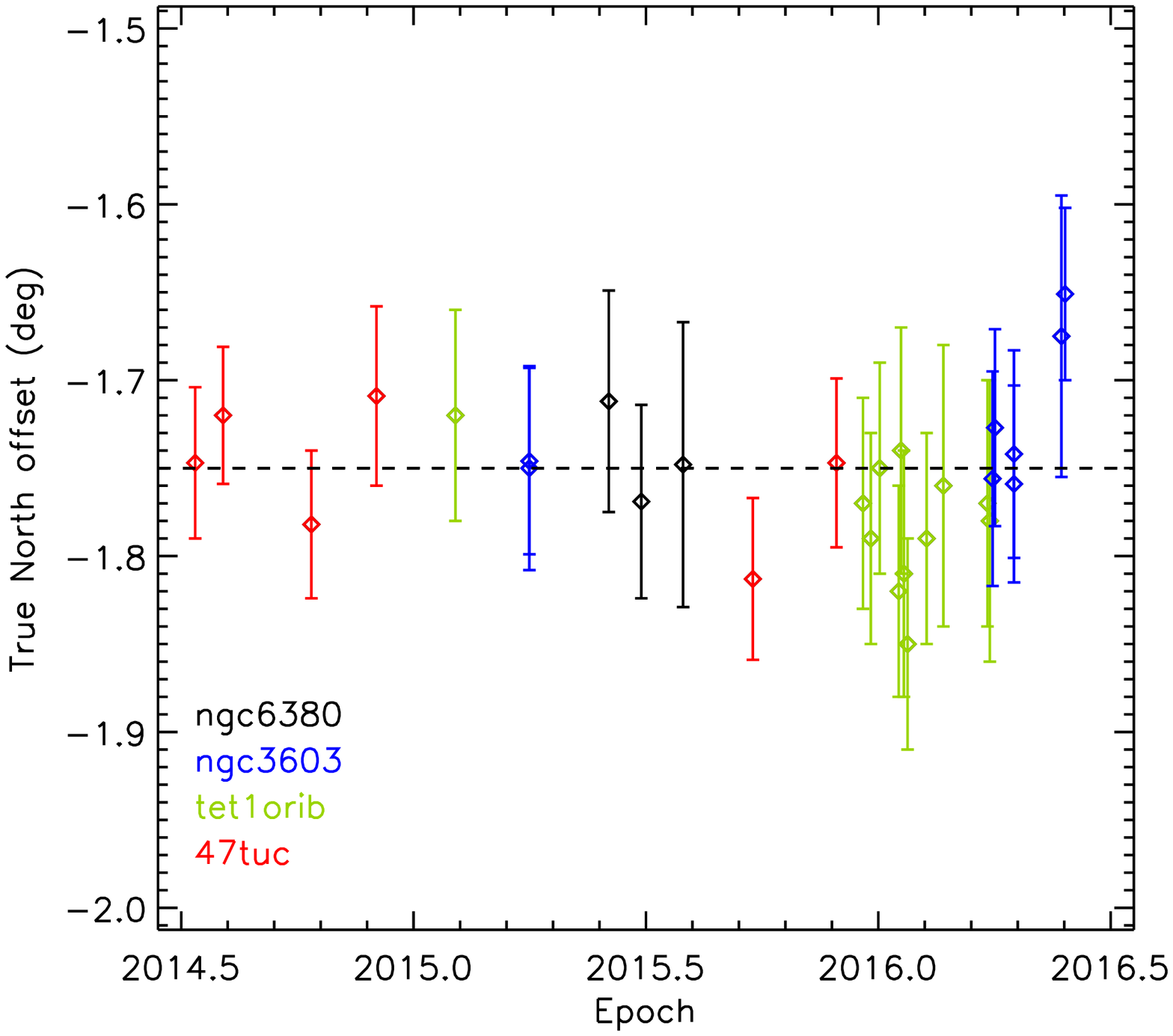}
   \end{tabular}
   \end{center}
   \caption[fields] 
   { \label{fig:irdissumaryplots} 
Individual measurements for the pixel scale (left) and true North (right) of IRDIS. For the pixel scale measurements, only the configuration H2 filter + N\_ALC\_YJH\_S coronagraph is considered.\\} 
   \end{figure}

After applying this correction to the true North data, we obtained the measurements listed in Tab.~\ref{tab:irdistn} and displayed in the right panel of Fig.~\ref{fig:irdissumaryplots}. The most accurate measurements (with uncertainties below $\sim$0.08$^{\circ}$) show small variations of $\sim$0.15$^{\circ}$ with an average value of $-$1.75$^{\circ}$. For a given stellar cluster, the error bars can vary from one observation to another because of the sensitivity (integration time, use or not of a coronagraph) and the quality of the images.

\begin{table}[ht]
\caption{Individual measurements for the IRDIS true North offset as a function of the observing run.} 
\label{tab:irdistn}
\begin{center}       
\begin{tabular}{|l|l|c|l|} 
\hline
\rule[-1ex]{0pt}{3.5ex}  Date & Field & True North ($^{\circ}$) \\ 
\hline
\rule[-1ex]{0pt}{3.5ex}  2014-07-18 & 47 Tuc & $-$1.747$\pm$0.043 \\ 
 \hline
 \rule[-1ex]{0pt}{3.5ex}  2014-08-05 & 47 Tuc & $-$1.720$\pm$0.039 \\ 
 \hline
 \rule[-1ex]{0pt}{3.5ex}  2014-10-11 & 47 Tuc & $-$1.782$\pm$0.042 \\ 
 \hline
 \rule[-1ex]{0pt}{3.5ex}  2014-12-02 & 47 Tuc & $-$1.709$\pm$0.051 \\ 
 \hline
 \rule[-1ex]{0pt}{3.5ex}  2015-02-03 & $\Theta^1$ Ori B & $-$1.72$\pm$0.06 \\ 
 \hline
 \rule[-1ex]{0pt}{3.5ex}  2015-03-31 & NGC 3603 & $-$1.750$\pm$0.058 \\ 
 \hline
 \rule[-1ex]{0pt}{3.5ex}  2015-03-31 & NGC 3603 & $-$1.746$\pm$0.053 \\ 
 \hline
  \rule[-1ex]{0pt}{3.5ex}  2015-05-03 & HIP 67745 & $-$1.86$\pm$0.15 \\ 
 \hline
 \rule[-1ex]{0pt}{3.5ex}  2015-05-03 & HIP 68725 & $-$2.33$\pm$0.14 \\ 
 \hline 
 \rule[-1ex]{0pt}{3.5ex}  2015-05-30 & NGC 6380 & $-$1.712$\pm$0.063 \\ 
 \hline
 \rule[-1ex]{0pt}{3.5ex}  2015-06-28 & NGC 6380 & $-$1.769$\pm$0.055 \\ 
 \hline
 \rule[-1ex]{0pt}{3.5ex}  2015-07-31 & NGC 6380 & $-$1.748$\pm$0.081 \\ 
 \hline
 \rule[-1ex]{0pt}{3.5ex}  2015-09-24 & 47 Tuc & $-$1.813$\pm$0.046 \\ 
 \hline
 \rule[-1ex]{0pt}{3.5ex}  2015-11-29 & 47 Tuc & $-$1.747$\pm$0.048 \\ 
 \hline
 \rule[-1ex]{0pt}{3.5ex}  2015-12-20 & $\Theta^1$ Ori B & $-$1.77$\pm$0.06 \\ 
 \hline
  \rule[-1ex]{0pt}{3.5ex}  2015-12-26 & $\Theta^1$ Ori B & $-$1.79$\pm$0.06 \\ 
 \hline
\end{tabular}
\hspace{0.6cm}
\begin{tabular}{|l|l|c|l|} 
\hline
\rule[-1ex]{0pt}{3.5ex}  Date & Field & True North ($^{\circ}$) \\ 
 \hline
 \rule[-1ex]{0pt}{3.5ex}  2016-01-02 & $\Theta^1$ Ori B & $-$1.75$\pm$0.06 \\ 
 \hline
  \rule[-1ex]{0pt}{3.5ex}  2016-01-16 & $\Theta^1$ Ori B & $-$1.82$\pm$0.06 \\ 
 \hline
  \rule[-1ex]{0pt}{3.5ex}  2016-01-18 & $\Theta^1$ Ori B & $-$1.74$\pm$0.07 \\ 
 \hline
 \rule[-1ex]{0pt}{3.5ex}  2016-01-20 & $\Theta^1$ Ori B & $-$1.81$\pm$0.07 \\ 
 \hline
 \rule[-1ex]{0pt}{3.5ex}  2016-01-23 & $\Theta^1$ Ori B & $-$1.85$\pm$0.06 \\ 
 \hline
 \rule[-1ex]{0pt}{3.5ex}  2016-02-07 & $\Theta^1$ Ori B & $-$1.79$\pm$0.06 \\ 
 \hline
 \rule[-1ex]{0pt}{3.5ex}  2016-02-20 & $\Theta^1$ Ori B & $-$1.76$\pm$0.08 \\ 
 \hline
  \rule[-1ex]{0pt}{3.5ex}  2016-03-26 & $\Theta^1$ Ori B & $-$1.77$\pm$0.07 \\ 
 \hline
 \rule[-1ex]{0pt}{3.5ex}  2016-03-28 & $\Theta^1$ Ori B & $-$1.78$\pm$0.08 \\ 
 \hline
 \rule[-1ex]{0pt}{3.5ex}  2016-03-30 & NGC 3603 & $-$1.756$\pm$0.061 \\ 
 \hline
 \rule[-1ex]{0pt}{3.5ex}  2016-04-01 & NGC 3603 & $-$1.727$\pm$0.056 \\ 
 \hline
 \rule[-1ex]{0pt}{3.5ex}  2016-04-16 & NGC 3603 & $-$1.742$\pm$0.059 \\ 
 \hline
  \rule[-1ex]{0pt}{3.5ex}  2016-04-16 & NGC 3603 & $-$1.759$\pm$0.056 \\ 
 \hline
  \rule[-1ex]{0pt}{3.5ex}  2016-05-22 & NGC 3603 & $-$1.675$\pm$0.080 \\ 
 \hline
  \rule[-1ex]{0pt}{3.5ex}  2016-05-25 & NGC 3603 & $-$1.651$\pm$0.049 \\ 
 \hline
\end{tabular}
\end{center}
\end{table}

A component was installed in the SPHERE LCU in July 2016 for ensuring its proper synchronization with the telescope internal clock. Dome and on-sky tests are on-going for checking that the missynchronization issues are effectively solved.

\subsection{IFS}
\label{sec:ifs}

\subsubsection{Pixel scale of the reconstructed data cubes}
\begin{figure}[t]
   \begin{center}
   \begin{tabular}{c} 
	\includegraphics[height=5.8cm]{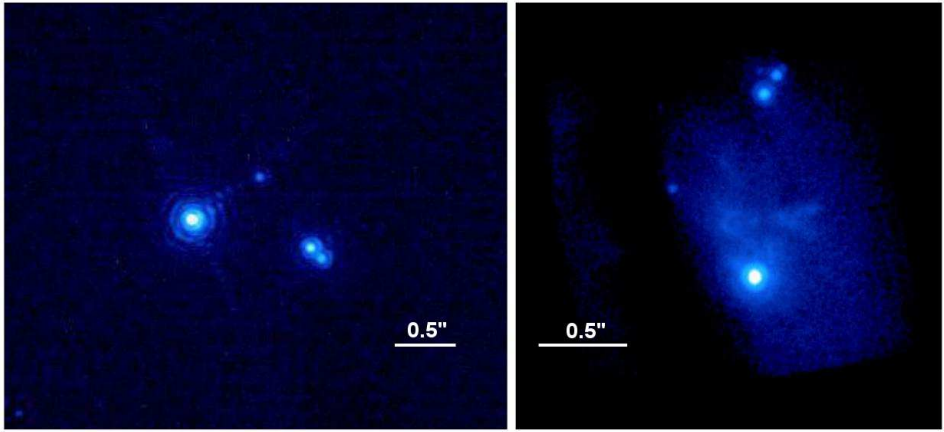}
   \end{tabular}
   \end{center}
   \caption[fields] 
   { \label{fig:irdisifspup} 
IRDIS (left) and IFS (right) images of the Orion Trapezium B1--B4 quadruple.} 
   \end{figure}

Due to the small field of view of IFS, direct on-sky astrometric calibration is currently feasible on the Orion Trapezium B1--B4 quadruple only (Fig.~\ref{fig:irdisifspup}). The measurements give values of the pixel scale of 7.46$\pm$0.02~mas/pixel (specifications 7.4$\pm$0.1~mas/pix), which is the value used as default in the SPHERE Data Center pipeline. Nevertheless, since IFS can be operated only in parallel with IRDIS, indirect calibrations are derived using the on-sky IRDIS pixel scale and the IRDIS/IFS pixel scale ratio from IRDIS and IFS distortion grid data recorded simultaneously (Fig.~\ref{fig:distortiongrids}). 

\subsubsection{Angle offset between the IFS and IRDIS fields of view}
Since IFS operates in parallel with IRDIS, the IRDIS true North measurement from any run (Tab.~\ref{tab:irdistn}) is also applied to IFS images. An additional derotation is included to align the North up and the East to the left since the IFS field of view is rotated with respect to the IRDIS field of view (Fig.~\ref{fig:irdisifspup}). The current estimate of this relative orientation is +100.48$\pm$0.10$^{\circ}$. The on-sky estimate is consistent with measurements obtained using internal calibration grids of the two instruments (+100.55$\pm$0.10$^{\circ}$).

\section{Summary and prospects}
\label{sec:conclusions}
We have presented the current on-sky results for the astrometric calibration of the SPHERE near-infrared instruments (IRDIS and IFS) based on data collected over 2~years of operations. On-sky measurements of the SPHERE+VLT optical distortion show that the optical distortion from the VLT is negligible with respect to the SPHERE optical distortion and that the main SPHERE distortion effect results in a horizontal pixel scale 0.60$\pm$0.02\% larger than the vertical pixel scale for the IRDIS camera. Then, we estimated the zeropoint angle of the SPHERE pupil in pupil-stabilized mode, which is the main observing mode used for the consortium guaranteed-time exoplanet imaging survey.  This parameter is stable around an average value of $-$135.99$\pm$0.11$^{\circ}$. We measured the pixel scale for IRDIS for various filter pairs and coronagraphs as well as the true North offset from July 2014 to May 2016. For given instrument configuration and astrometric calibration field, the pixel scale has a stability over a timescale of a few months of 0.009~mas/pixel. For coronagraphic data obtained with the H2 filter, our current estimate of the pixel scale is 12.255$\pm$0.009~mas/pix. Technical tracking tests with the SPHERE internal distortion grid showed missynchronization issues between the SPHERE and VLT internal clocks. These missynchronization issues produced anomalous variations larger than 1$^{\circ}$ for the true North measurements between December 2015 and February 2016, but these issues existed since the commissioning phase. The true North deviations are clearly correlated with the derotation error produced by the clock missynchronization. The latter can be derived from the angle information in the data headers. After correcting the true North measurements for the derotation error, the statistics give a remarkable stable value of $-$1.75$\pm$0.08$^{\circ}$. For the IFS reconstructed data cubes, on-sky and internal calibration data indicate an average pixel scale value of 7.46$\pm$0.02~mas/pix and a relative orientation to the IRDIS field of view of +100.48$\pm$0.10$^{\circ}$.

As part of the consortium guaranteed-time observations, we will continue to monitor the SPHERE astrometric parameters for the full survey. The public release of the GAIA data in late 2016 will provide absolute astrometric calibration for several of the SPHERE astrometric fields, which will help in improving the absolute calibration of the SPHERE data. We plan to release our consortium astrometric tool and observing procedures to ESO so that such analyses can also be carried out on a regular basis by the observatory. The results from these analyses will be made available to the SPHERE users outside the instrument consortium on a public web page.

Common astrometric fields observed with various high-contrast imaging instruments (e.g., SPHERE, GPI, LMIRCam, MagAO, CHARIS) will allow for a better comparison of astrometric measurements from different instruments and a reduction of the systematic errors, which are a major issue for the determination of the orbital properties of directly-imaged companions in particular for those located at wide separations. We note that the Orion Trapezium field is very suitable to these comparisons because of its observability from both northern and southern hemispheres.

\acknowledgments 
We thank Andrea Bellini, Jay Anderson, Eva Noyola, and Zeinab Khorrami for kindly providing the HST positions of 47~Tucanae, NGC~6380, and NGC~3603, as well as Brian Mason for providing the astrometric data for the binaries. We also thank the ESO Paranal staff for support for conducting the observations reported in this paper. This research has made use of the Washington Double Star Catalog maintained at the U.S. Naval Observatory. SPHERE is an instrument designed and built by a consortium consisting of IPAG (Grenoble, France), MPIA (Heidelberg, Germany), LAM (Marseille, France), LESIA (Paris, France), Laboratoire Lagrange (Nice, France), INAF-Osservatorio di Padova (Italy), Observatoire de Gen\`eve (Switzerland), ETH Zurich (Switzerland), NOVA (Netherlands), ONERA (France), and ASTRON (Netherlands), in collaboration with ESO. SPHERE was funded by ESO, with additional contributions from the CNRS (France), MPIA (Germany), INAF (Italy),
FINES (Switzerland), and NOVA (Netherlands). SPHERE also received funding from the European Commission Sixth and Seventh Framework Programs as part of the Optical Infrared Coordination Network for Astronomy (OPTICON) under grant number RII3-Ct-2004-001566 for FP6 (2004--2008), grant number 226604 for FP7 (2009--2012), and grant number 312430 for FP7 (2013--2016). A.-L.M., R.G., S.D., and R.U.C acknowledge support from the ``Progetti Premiali'' funding scheme of the Italian Ministry of Education, University, and Research.

\bibliography{./biblio} 
\bibliographystyle{spiebib} 

\end{document}